\documentclass[12pt,preprint]{aastex}

\newcommand{\etal  }{{et al.} }

\def\lesssim{\mathrel{\hbox{\rlap{\hbox{\lower4pt\hbox{$\sim$}}}\hbox{$<$}}}}
\def\gtrsim{\mathrel{\hbox{\rlap{\hbox{\lower4pt\hbox{$\sim$}}}\hbox{$>$}}}}
\newcommand{\cm  }{\,{\rm cm}^{-3} } 
\newcommand{\nc  }{n_{\rm c} } 
\newcommand{\zsun  }{Z_{\odot} }

\shorttitle{Binary Formation}
\shortauthors{Machida  2008}

\begin{document}

\title{Binary Formation in Star-Forming Clouds with Various Metallicities}

\author{Masahiro N. Machida\altaffilmark{1}} 

\altaffiltext{1}{Department of Physics, Graduate School of Science, Kyoto University, Sakyo-ku, Kyoto 606-8502, Japan; machidam@scphys.kyoto-u.ac.jp}

\begin{abstract}
Cloud evolution for various metallicities is investigated by three-dimensional nested grid simulations, in which the initial ratio of rotational to gravitational energy of the host cloud $\beta_0$ ($=10^{-1}$--$10^{-6}$) and cloud metallicity $Z$ ($=0$--$\zsun$) are parameters.
Starting from a central number density of $\nc = 10^4 \cm$, cloud evolution for 48 models is calculated until the protostar is formed ($\nc \simeq10^{23}\cm$) or fragmentation occurs.
The fragmentation condition depends both on the initial rotational energy and cloud metallicity.
Cloud rotation promotes fragmentation, while fragmentation tends to be suppressed in clouds with higher metallicity.
Fragmentation occurs when $\beta_0>10^{-3}$ in clouds with solar metallicity ($Z=\zsun$), while fragmentation occurs when $\beta_0>10^{-5}$ in the primordial gas cloud ($Z=0$).
Clouds with lower metallicity have larger probability of fragmentation, which indicates that the binary frequency is a decreasing function of cloud metallicity.
Thus, the binary frequency at the early universe (or lower metallicity environment) is higher than at present day (or higher metallicity environment).
In addition, binary stars born from low-metallicity clouds have shorter orbital periods than those from high-metallicity clouds. 
These trends are explained in terms of the thermal history of the collapsing cloud.

\end{abstract}
\keywords{binaries: general---cosmology: theory---early universe---stars: formation}

\section{Introduction}
Observations have shown that about 60--80\% of field stars are members of binary or multiple systems \citep[e.g.,][]{abt83,duq91}.
Hence, a majority of main-sequence stars belong to binary or multiple systems.
These stars have a metallicity equivalent to that of the sun (i.e., solar metallicity).
On the other hand, \citet{lucatello05} investigated radial velocities of carbon-enhanced, $s$-process-rich, very metal-poor (CEMP-s) stars that constitute a substantial proportion of the extremely metal-poor (EMP) stars of the Galactic halo.
In their observations, they found that about 100\% of these metal-poor stars are in binary (or multiple) systems, and the orbital periods of these stars are shorter than those of field stars in the solar neighborhood.
Recently, EMP stars with [Fe/H] $<-5$ (HE0107-5240 and HE1327-2326) have been observed \citep{christlieb01,frebel05}.
\citet{suda04} showed that the observed pattern of metal abundance for these stars can be explained by nucleosynthesis and mass transfer in a first-generation low-mass binary star.
\citet{komiya07} investigated the origin of EMP stars in the context of stellar evolution and found that the observed EMP stars were exclusively born as secondary members of binaries.
Thus, these studies imply a high binary frequency in the early universe (or low metallicity environment).

For the present day star formation process, detailed numerical simulations studies of fragmentation and binary formation process have been performed by many authors \citep[see][]{bodenheimer00,goodwin07}. 
The cloud evolution has been calculated from $\nc \simeq 10^4\cm$ to $\simeq 10^{21}\cm$, where $\nc$ is the number density at the cloud center. The studies have shown that fragmentation occurs only for $10^{11}\cm \lesssim \nc \lesssim 10^{15}\cm$ \citep[see also][]{bate98,whitehouse06}.

On the other hand, in the collapsing primordial cloud ($Z=0$), fragmentation rarely occurs for $10^4\lesssim \nc \lesssim10^{16}\cm$ \citep[e.g.,][]{yoshida06,bromm02,abel02} and frequently occurs for $n\gtrsim10^{16}\cm$ \citep{saigo04, machida08b}.
\citet{machida08b} also showed that binary separation in the early universe is narrower than that at the present day.
Recently, star formation in a collapsing low metallicity ($Z<10^{-4}\zsun$) gas cloud has been studied by \citet{smith07} and \citet{clark08}.
\citet{clark08} found that binary or multiple stellar systems can form even in a low metallicity cloud.
However, in their studies, since they adopted a sink cell, they calculated the cloud evolution only up to $\nc \simeq 10^{16}\cm$.
Fragmentation may occur in a later collapsing phase ($\nc \gtrsim 10^{16}\cm$), as shown in \citet{machida08b}.

In this paper, we adopt a barotropic equation of state and calculate the evolutions of collapsing clouds with various metallicities ($Z=0-\zsun$), from the formation of dense cores ($n\simeq 10^4\cm$)  up to stellar core formation ($n\simeq 10^{23}\cm$).
The calculations indicate that the binary frequency increases as cloud metallicity lowers, and binary separation in lower metallicity clouds is narrower than in higher metallicity clouds.

\section{Model and Numerical Method}
 To study the evolution of star-forming cores in a large dynamic range of density and spatial scale,  a three-dimensional nested grid method is used
and the equations of hydrodynamics including self-gravity are solved (see Eq. [1]--[3] of \citealt{machida08b}).  
For gas pressures in collapsing clouds with different metallicities, barotropic relations that approximate the results of \citet{omukai05} are adopted.
Figure~\ref{fig:1} shows these relations (thick solid lines) as well as the data from \citet[][thick dotted line]{omukai05} for pressure evolution with different metallicities plotted as a function of number density . 
To stress the variations of pressure with density, $P/n$ is plotted \citep[for details see][]{omukai05}, which is proportional to the gas temperature if the mean molecular weight is constant.

As the initial state we take a spherical cloud of density $1.4$ times higher than the hydrostatic equilibrium with external pressure (i.e., the so-called critical Bonnor--Ebert sphere).
The initial central density is taken as $n_{\rm c,0}$ = $1.4\times10^4\cm $.
Each cloud rotates rigidly ($\Omega_0$) around the $z$-axis.
The initial temperatures, which are derived from \citet{omukai05}, are different in clouds with different metallicities (see dotted lines of Fig.~1).
For example, a cloud with $Z=0$ (primordial composition) has an initial temperature of 230\,K, while a cloud with $Z=\zsun$ (solar composition) has 7\,K as the initial temperature. 
Since critical Bonnor--Ebert spheres are assumed as the initial state, the radii of the initial spheres are different depending on the initial temperature (or assumed metallicity): the radius of a $Z=0$ cloud is $5.5\times10^5$\,AU, while that of a $Z=\zsun$ cloud is $1.2\times 10^5$\,AU.
We have confirmed that these initial differences of radius do not greatly affect the subsequent cloud evolution in calculations in which the size and density of the initial cloud are changed.

The models are characterized by two parameters: the initial rotational energy $\beta_0$ and the cloud metallicity $Z$. The values used for these parameters are  $\beta_0$  = $10^{-1}$, $10^{-2}$, $10^{-3}$, $10^{-4}$, $10^{-5}$ and $10^{-6}$,  and $Z$ = $0$, $10^{-6}$, $10^{-5}$, $10^{-4}$, $10^{-3}$, $10^{-2}$, $10^{-1}$, and $1\zsun$.
Combining these two ranges of values, cloud evolution for 48 models is investigated. 
To induce fragmentation, 1\% of the non-axisymmetric density perturbation of the $m=2$ mode (i.e., bar mode) is added to the initial cloud.

To calculate a large spatial scale, the nested grid method is adopted \citep[for details see ][]{matsumoto04,machida05a,machida05b}. 
Each level of a rectangular grid has the same number of cells ($ 128 \times 128 \times 8 $), although the cell width $h(l)$ depends on the grid level $l$.
The cell width is reduced by a factor of two for every upper level.
The calculation is first performed with three grid levels ($l=1,2,3$).
The box size of the coarsest grid $l=1$ is chosen to be $2 R_{\rm c}$, where $R_{\rm c}$ is the radius of the critical Bonnor--Ebert sphere.
A new finer grid is generated whenever the minimum local Jeans length $ \lambda _{\rm J} $ becomes smaller than $ 8\, h (l_{\rm max}) $, where $h$ is the cell width.
The maximum level of grids is restricted to $l_{\rm max} \leqq 30$.

\section{Results}
 In each model, starting from a nearly hydrostatic core with central density $\nc = 10^4 \cm$, the evolution of the collapsing cloud is calculated.
When fragmentation does not occur, the cloud evolution is calculated until the protostar is formed at $\nc \simeq 10^{22}\cm$.
When fragmentation does occur,  the calculation is often stopped after fragmentation (i.e., before the protostar formation), because fragments escape from the finest grid.

Figure~\ref{fig:2} shows the final state for each model against the metallicity $Z$ ($x$-axis) and initial rotation energy $\beta_0$ ($y$-axis).
In the figure, the cloud evolutions are classified into four types: fragmentation (red panel border), merger (violet), non-fragmentation (blue), and stable core (gray) models.
Fragmentation occurs and two or more fragments appear in fragmentation and merger models.
After fragmentation, fragments survive without merger until the end of the calculation in fragmentation models, while fragments merge to form a single core in merger models.
In non-fragmentation and stable core models, fragmentation does not occur.
A single protostar is formed in non-fragmentation models, while a long-lived core is formed before protostar formation in stable core models.

\begin{figure}
\includegraphics[width=150mm]{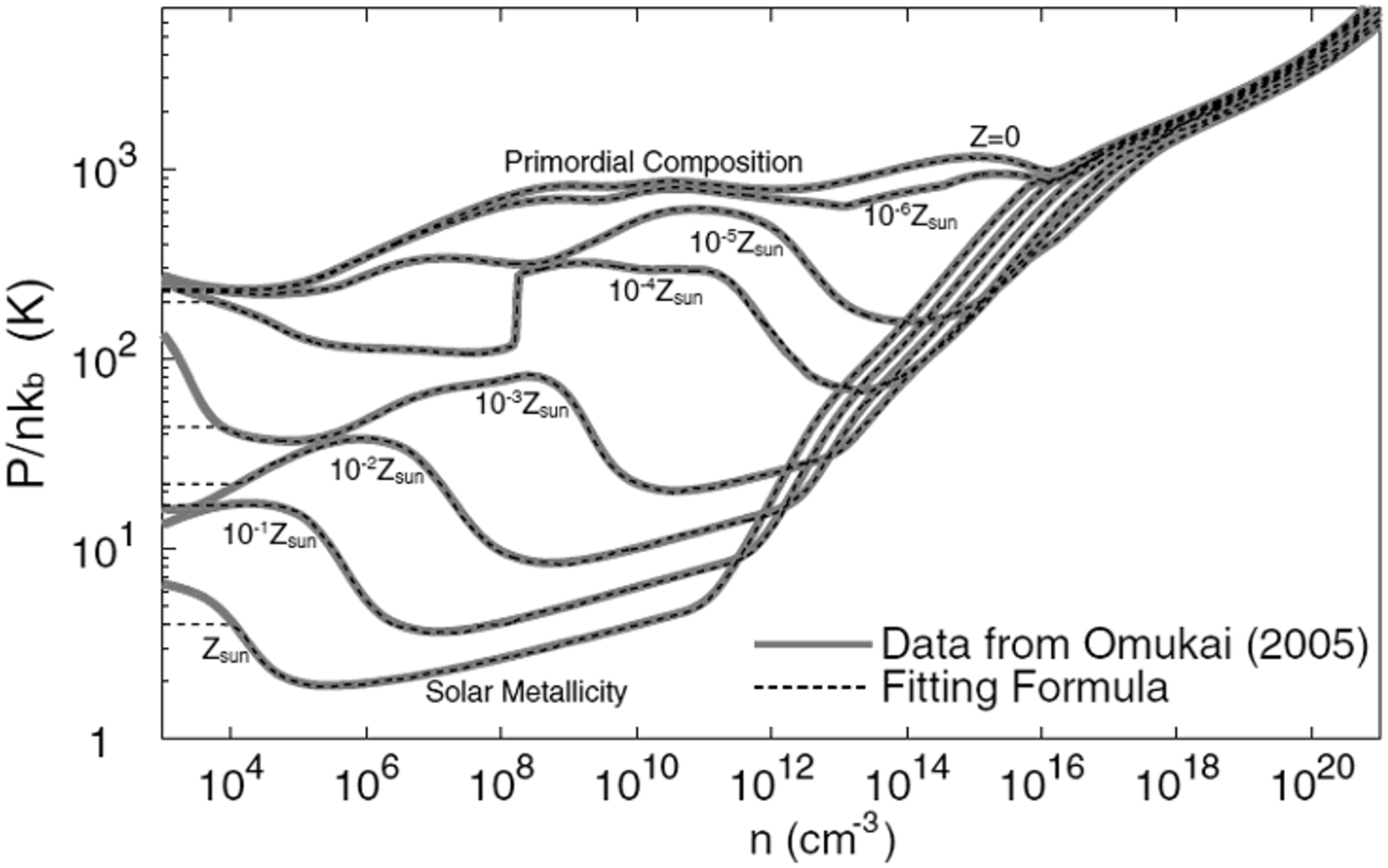}
\caption{
Thermal evolution of collapsing clouds against gas number density with different metallicities ($Z = 0$, $10^{-6}$, $10^{-5}$, $10^{-4}$, $10^{-3}$, $10^{-2}$, $10^{-1}$, $1\,\zsun$).
To stress the variation of pressure with density, $P/n$ is plotted, which is proportional to the gas temperature if the mean molecular weight is constant.
The thick solid lines show data from \citet{omukai05}.
The thick dotted lines are fits by our numerical simulation.
}
\label{fig:1}
\end{figure}

The rightmost column in Figure~\ref{fig:2} shows the final states for models with solar metallicity ($Z=\zsun$). 
Including radiative effects, the evolution of clouds with solar metallicity has been investigated by many authors. 
Assuming spherical symmetry, the evolution from the molecular cloud to the stellar core has been calculated by many authors \citep[e.g.,][]{larson69,masunaga00}.
In three dimensions, \citet{whitehouse06} and \citet{stamatellos07} have calculated stellar core formation from the molecular cloud core.
These studies have shown that the molecular gas obeys the isothermal equation of state with a temperature of $\sim 10$ K until $n_c \simeq 10^{11}\cm$, then the cloud collapses almost adiabatically  ($ 10^{11}\cm \lesssim n_c \lesssim 10^{16}\cm$; adiabatic phase) and a quasi-static core  (hereafter, first adiabatic core) forms during the adiabatic phase. 
Thermal evolution for a cloud with solar metallicity is also confirmed in Figure~\ref{fig:1}, in which the gas temperature increases adiabatically after the number density reaches $n\simeq 10^{11}\cm$.
In models with solar metallicity, the first adiabatic core surrounded by a shock front is formed at $\nc=10^{11}-10^{14}\cm$.

The rightmost column of Figure~\ref{fig:2} (models with $Z=\zsun$) shows that fragmentation occurs when the initial cloud has $\beta_0>10^{-3}$.
In these clouds, fragmentation occurs only after the first adiabatic core formation:  fragmentation occurred at $\nc=6.8\times 10^{11}\cm$ ($\beta_0=10^{-1}$), $\nc =2.4\times 10^{12}\cm$ ($\beta_0=10^{-2}$), and $\nc=1.7\times 10^{13}\cm$ ($\beta_0=10^{-3}$), respectively.
In these fragmentation models, fragments survived without merger until the end of the calculation for models with $\beta_0=10^{-1}$ and $10^{-2}$, while a single stable adiabatic core was formed after merger in the model with $\beta_0=10^{-3}$.
The fragmentation condition ($\beta_0 > 10^{-3}$) for a cloud with solar metallicity is consistent with that of \citet{matsu03}.
\citet{matsu03} calculated the evolution of a cloud with solar metallicity in a large parameter space and found that fragmentation occurs when the initial cloud has $\beta_0 > 2.2\times10^{-3}$.
In addition, many studies have shown that fragmentation occurs only after the first adiabatic core formation \citep[see][]{goodwin07}.

For clouds with $Z=\zsun$, fragmentation occurs in models with $\beta_0 \ge 10^{-3}$ after the first adiabatic core formation, while stable first adiabatic cores are formed in models with $\beta_0<10^{-4}$.
In models with $\beta_0 < 10^{-4}$, although the cloud evolution was calculated for a sufficiently long time, the first adiabatic core did not collapse to reach the protostellar density ($\nc \simeq 10^{22}\cm$).
When a non-rotating cloud is adopted as the initial state, the first adiabatic core increases its mass with time by gas accretion, and can collapse again (the second collapse, see \citealt{masunaga00}) to form a protostar for a short duration.
On the other hand, a long-lived first adiabatic core is frequently formed in a rotating cloud, as shown in \citet{saigo06}.
This long-lived core can collapse further after the non-axisymmetric perturbation grows sufficiently owing to the bar mode instability \citep{durisen86}, because the angular momentum of the core is transferred by the non-axisymmetric structure \citep{bate98,saigo06}.
To investigate further the evolution for models with $\beta_0 < 10^{-4}$,  in different models of Figure~\ref{fig:2}, a large initial amplitude of the non-axisymmetric perturbation is adopted.
In these calculations,  after the first adiabatic core formation, the non-axisymmetric pattern grows and the cloud collapses (the second collapse) to reach the stellar density ($\nc \simeq 10^{22}\cm$).
However, these models do not show fragmentation even in the later evolution phase ($\nc \gtrsim 10^{14}\cm$), as shown in \citet{bate98}.
Thus, it is expected that fragmentation does not occur in stable core models even after a calculation for an extended time period.
This is because  the central region transforms from a disk-like to a spherical configuration due to the large thermal energy after the first adiabatic core formation; fragmentation can easily occur in a thin disk-like configuration.
In addition, the angular momentum is effectively transferred from the high-density region by the non-axisymmetric pattern.

The leftmost column of Figure~\ref{fig:2} shows final states for models with $Z=0$ (primordial cloud).
In the primordial cloud, fragmentation occurs in clouds with $\beta_0>10^{-6}$, which is consistent with the results of \citet{machida08b}.
These panels also show that the fragmentation scale (i.e., distance between fragments) increases with $\beta_0$.
As shown in \citet{machida08a, machida08b}, the fragmentation scale is comparable to the Jeans length at the fragmentation epoch.
Since fragmentation occurs in the earlier epoch (or lower density) for models with larger $\beta_0$ and the Jeans length shortens as the density increases, fragments have a larger separation in clouds with larger $\beta_0$.

The first adiabatic core is formed at $\nc=10^{11}$--$10^{14}\cm$ in clouds with solar metallicity, while the protostar is formed directly without the first adiabatic core formation in the primordial cloud.
As shown in Figure~\ref{fig:1}, since the temperature in the primordial cloud ($Z=0$) gradually increases with $\gamma\simeq 1.1$ of the polytropic index \citep{omukai98} in a wide density range of $10^3\cm<\nc<10^{18}\cm$, the (first) adiabatic core is not formed before the protostar formation ($\nc\lesssim10^{21}\cm$).
Thus, fragmentation can occur for  $\nc \lesssim 10^{21}\cm$ in the primordial cloud, while fragmentation occurs only for $\nc \lesssim 10^{14}\cm$ in clouds with solar metallicity.
In addition, in the primordial cloud, since the central region can be spun until the protostar is formed ($\nc \lesssim 10^{21}\cm$) due to the absence of the first adiabatic core \citep[see][]{machida07}, fragmentation occurs even in clouds with smaller $\beta_0$.

In Figure~\ref{fig:2}, the border between fragmentation and non-fragmentation models is shown by a gray-white dotted line, which indicates that fragmentation tends to occur in clouds with lower metallicity.
For example, fragmentation occurs in models with $\beta_0 > 10^{-4}$ for clouds with $Z$ = $10^{-3}\zsun$ and $10^{-4}\zsun$, while fragmentation occurs for $\beta_0 > 10^{-3}$ for clouds with $Z=10^{-1}\zsun$ and $10^{-2}\zsun$.

The fragmentation epoch is closely related to the first adiabatic core formation epoch.
The first adiabatic core is formed in clouds with $Z\ge10^{-5}\zsun$, while the first adiabatic core is not formed in clouds with $Z<10^{-5}\zsun$.
As shown in Figure~\ref{fig:1}, the gas temperature increases keeping $\gamma\simeq 1.1$ for clouds with $Z=0$ and $10^{-6}\zsun$, while the gas temperature increases adiabatically ($\gamma\simeq 1.4$) after the central density reaches $\nc \simeq 10^{11}$--$10^{16}\cm$ in clouds with $Z\le10^{-5}\zsun$.
In clouds with $Z \le 10^{-5}\zsun$, the epoch when the cloud collapses adiabatically depends on the cloud metallicity. The first adiabatic core is formed at an earlier epoch (or lower density) in clouds with lower metallicity.
For example, the first adiabatic core is formed at $\nc \simeq 5\times 10^{11}\cm$ in clouds with ($Z$, $\beta_0$) = ($10^{-1}$, $10^{-6}$), while the first adiabatic core is formed at $\nc \simeq 6\times 10^{14}\cm$ in clouds with ($Z$, $\beta_0$) = ($10^{-5}$, $10^{-6}$).
When the first adiabatic core is formed at a later evolution phase (or higher density), the cloud is spun over a long duration and forms a thin rotating disk even in clouds with small $\beta_0$.
Then, fragmentation occurs in the thin disk.

\section{Discussion}
\begin{figure}
\plotone{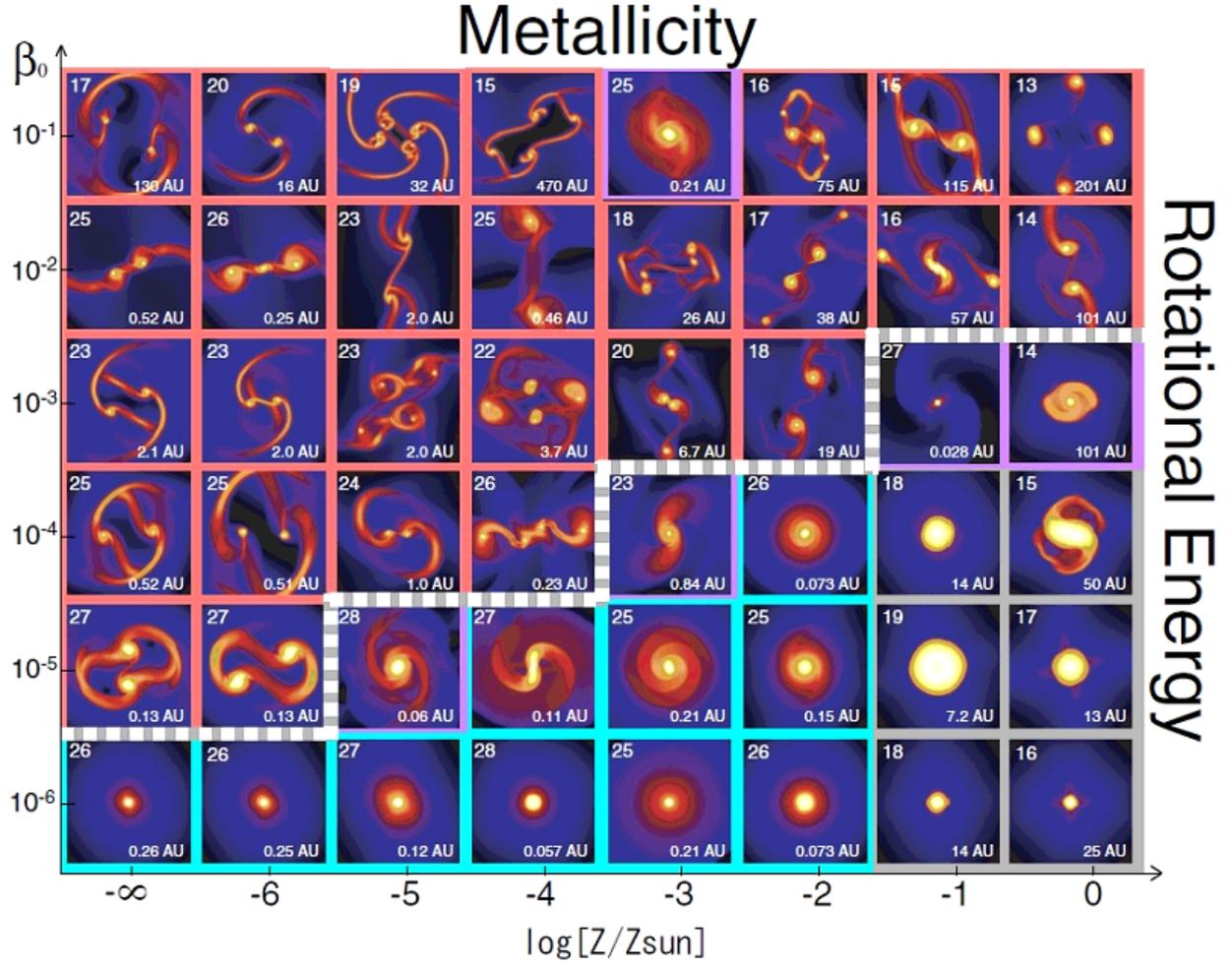}
\caption{
Outcomes of the collapse of prestellar clouds against different metallicities $Z$ and initial rotation energies $\beta_0$.
The density distribution (color-scale) around the center on the equatorial plane is shown in each panel.
The grid level is displayed at the upper left corner of each panel
and the grid scale is denoted at each lower right corner.
The panel border color indicates the following: red: fragmentation model, violet: merger model, blue: non-fragmentation model, and gray: stable core model.
The gray-white dotted line shows the border between fragmentation and non-fragmentation models.
}
\label{fig:2}
\end{figure}

Although we cannot observe the primordial gas cloud in the early universe, we have observed many molecular clouds in the solar neighborhood.
\citet{caselli02} observed about 60 molecular cloud cores and found them to have rotational energies in the range $\beta_0 = 10^{-4}$--$0.07$, with a typical value of $\beta_0=0.02$ (see also \citealt{goodman93}).
Figure~\ref{fig:2} shows that fragmentation occurs in clouds with solar metallicity ($Z=\zsun$) when the rotational energy exceeds $\beta_0>10^{-3}$, which is smaller than the typical value of the observations.
Thus, it is expected that fragmentation frequently occurs in clouds with solar metallicity.

Observations have shown a high binary frequency at the present day \citep[see][]{mathiu94}.
However, we cannot determine the binary frequency in the early universe (or binary frequency in a lower metal environment) from observations.
Cosmological simulations show that the first collapsed objects formed at $z\sim10$ with $\nc\simeq 10^3-10^4\cm$ have rotational energies comparable to or slightly larger than the typical value of molecular clouds \citep{bromm02,yoshida06}.
Thus, if the distribution of cloud rotational energy does not strongly depend on metal abundance, the binary frequency will increase with decreasing cloud metallicity.
Even if a typical low-metallicity cloud has a smaller rotational energy of $\beta_0 \ll 0.02$, the binary frequency may be high, because fragmentation can occur even in these clouds; fragmentation occurs in clouds  with $Z<10^{-4}\zsun$ when $\beta_0 \ge 10^{-4}$, which is much smaller than the typical value of molecular clouds ($\beta_0 = 0.02$).
Thus, it is expected that the binary frequency in low-metallicity clouds is higher than in solar-neighborhood clouds.

The fragmentation scale or separation between fragments decreases with decreasing cloud metallicity.
For example, at the fragmentation epoch, the separation between fragments for model ($Z$, $\beta_0$) = ($\zsun$, 0.1) is $\sim 45$\,AU, while the separation for ($Z$, $\beta_0$) = ($0$, 0.1) is $0.2$\,AU (see Fig.~\ref{fig:2}).
Since the first adiabatic core is formed at a higher density in clouds with lower metallicity, the fragmentation scale decreases with decreasing metallicity, i.e. the fragmentation scale (or Jeans scale) shortens as the cloud density increases.
Thus, at the moment of birth, the binary orbital period in low-metallicity clouds is shorter than that of solar-neighborhood clouds.

In the models shown in Figure~\ref{fig:2}, 1\% of the non-axisymmetric perturbation is added to the initial state.
When different initial amplitudes of the non-axisymmetric perturbation are added, the cloud evolution may change.
For example, in the fragmentation region of Figure~\ref{fig:2}, although fragments merge to form a single core only in model ($Z$, $\beta_0$) = ($10^{-3}$, 0.1), fragments do not merge in this model when different amplitudes of the non-axisymmetric perturbation (5, 10, 20, 30\%) are added.
\citet{goodwin07} pointed out that statistical studies are needed to understand fragmentation, because the cloud evolution in the high-density region is highly chaotic.
In subsequent papers, we will statistically investigate the cloud evolution for different initial cloud shapes in a large parameter space.

\acknowledgments
I thank ~K. Omukai for supplying the data of thermal evolution for the primordial collapsing cloud.
I also thank ~T. Matsumoto for contributing the nested grid code.
I have greatly benefited from discussions with ~S. Inutsuka.
Numerical computations were carried out on VPP5000 at the Center for Computational Astrophysics of the National Astronomical Observatory of Japan.
This work is supported by Grants-in-Aid from MEXT (18740104).

\clearpage

\end{document}